\documentclass[letter,twocolumn]{jpsj3}
\usepackage{txfonts}
\usepackage[dvipdfmx]{graphicx}
\usepackage{color}

\newcommand{\red}[1]{\textcolor{black}{#1}}

\title{Enhancement and Discontinuity of Effective Mass through the First-Order Metamagnetic Transition in UTe$_2$}

\author{
Atsushi Miyake$^1$\thanks{miyake@issp.u-tokyo.ac.jp}, 
Yusei Shimizu$^2$, 
Yoshiki J. Sato$^2$, 
Dexin Li$^2$, 
Ai Nakamura$^2$, 
Yoshiya Homma$^2$, 
Fuminori Honda$^{2, 3}$, 
Jacques Flouquet$^4$, 
Masashi Tokunaga$^1$, and 
Dai Aoki$^{2, 4}$
}

\inst{
$^1$The Institute for Solid State Physics, The University of Tokyo, Kashiwa, Chiba 277-8581, Japan\\
$^2$Institute for Materials Research, Tohoku University, Oarai, Ibaraki 311-1313, Japan \\
$^3$Central Institute of Radioisotope Science and Safety Management, Kyushu University, Fukuoka, 819-0395, Japan\\ 
$^4$University Grenoble Alpes, CEA, IRIG-PHELIQS, 38000 Grenoble, France
} 

\abst{
Metamagnetic transitions in the novel spin-triplet superconductor UTe$_2$ were investigated through the newly developed simultaneous measurements of magnetization and sample temperature for the field along the orthorhombic $b$-axis and close to the [011] direction, where reentrant superconductivity (RSC) is detected below and above the first-order metamagnetic transition field $H_{\rm m}$. 
Combining the Maxwell's relation and the Clausius-Clapeyron equation, we obtained the field dependence of Sommerfeld coefficient $\gamma$ through the first-order metamagnetic transition.
A significant enhancement of the effective mass toward $H_{\rm m}$ were detected for both field directions.
On the other hand, above $H_{\rm m}$, the effective mass discontinuously decreases for $H~||~b$, while it discontinuously increases for $H~||~\sim [011]$, which seems to make a crucial role for the RSC.}

\begin{document}
\maketitle

Recently discovered unconventional superconductivity of UTe$_2$ (space group: $Immm$) is a new promising candidate for the spin triplet state  \cite{Ran2019,Aoki2019}. 
The spin-triplet superconductivity is most likely realized in ferromagnetic (FM) systems, whose superconducting (SC) phase microscopically coexists with FM order \cite{Aok12_JPSJ_review, Aoki2019re}.
In this context, UGe$_2$ \cite{Sax00}, URhGe \cite{Aok01}, and UCoGe \cite{Huy07} are extensively studied.
For UTe$_2$, the SC upper critical fields $H_{c2}$ for any orthorhombic principal axes are far above the Pauli limitation \cite{Ran2019,Aoki2019}. 
Below the SC transition temperature, $T_{\rm sc}$, the claim of the spin-triplet pairing seems supported by the spin susceptibility data probed by NMR Knight shift experiments \cite{Nakamine2019, Nakamine2021}, a point node gap structure by specific heat, thermal conductivity and penetration depth \cite{Metz2019,Kittaka2020}, broken time-reversal symmetry by Kerr effect \cite{Hayes2020}, and chiral edge state through STM measurements\cite{Jiao2020}. 
Other important findings have been obtained through the pressure measurements. 
With increasing pressure, $T_{\rm sc}\sim 1.6$~K splits into two and the multiple SC phases appear \cite{Braithwaite2019, Ran2020, Aoki2020, Knebel2020,Thomas2020,Aoki2021}.

\begin{figure}
\begin{center}
\includegraphics[width=\hsize]{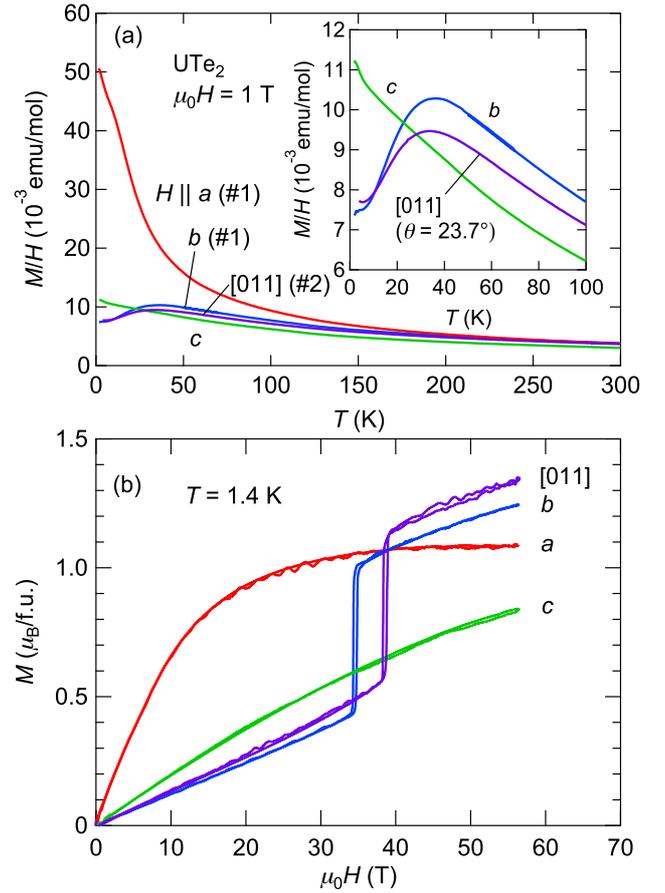}
\end{center}
\caption{(Color online)
(a)Temperature dependence of $M/H$ at 1~T for $H~||~a$, $b$ and $c$, and [011] axes of UTe$_2$.
The inset in panel (a) focuses near $T_{\chi}^{\rm max}$.
Magnetic field dependence of magnetization at different field directions along $a$, $b$, $c$, and [011] axes at 1.4 K.
The data for $H~||~a$, $b$, and $c$ in (a) and for $H~||~a$ and $b$ are taken from Ref \citen{Miyake2019}.
}
\label{MHMT}
\end{figure}

Magnetic field also induces the multiple SC phases in UTe$_2$.
In particular, $H$ along the $b$ axis induces quite non-trivial phenomena. 
$T_{\rm sc}$ decreases with field up to $\sim$15~T, but increases at higher fields up to the first-order metamagnetic transition (MMT) field $H_{\rm m}$ \cite{Ran2019b,Knebel2019,Niu2020, Lin2020,Knafo2021}.
Such a characteristic reentrant (R) SC phase diagram reminds us of the case of FM superconductors, URhGe and UCoGe.
While the SC phase in the URhGe and UCoGe is strongly suppressed by $H$ along the easy-magnetization $c$ axis, the transverse fields along the hard $b$ axis reemergence or reinforce of the SC phase \cite{Levy2005,Levy2007,Aoki2009}. 
In contrast to the FM superconductors, UTe$_2$ does not show any static magnetic order \cite{Sunder2019,Hutanu2020,Paulsen2021}, although FM fluctuations are suggested by NMR \cite{Tokunaga2019} and $\mu$SR \cite{Sunder2019}, while direct antiferromagnetic (AFM) correlations are detected by inelastic neutron scattering experiments \cite{Duan2020,Knafo2021b}.
Interestingly, this RSC phase of UTe$_2$ suddenly disappears accompanied by a first-order MMT at $\mu_0 H_m$~$\sim$~35~T \cite{Knebel2019, Ran2019b,Niu2020, Lin2020, Knafo2021}.
At $H_{\rm m}$, the electrical resistivity and magnetization jump discontinuously with the $H$ hysteresis  \cite{Knebel2019,Knafo2019,Miyake2019,Ran2019b,Lin2020,Knafo2021}.
Fermi surface (FS) reconstruction was also reported at $H_{\rm m}$\cite{Niu2020}.
$H_{\rm m}$ has similar energy scale to the maximum temperature ($T$) of magnetic susceptibility $\chi(T)$, $T_{\chi}^{\rm max}\sim$~35~K.
[see Fig.\ref{MHMT}(b)].
Above a critical end point (CEP) temperature, this MMT changes to a crossover, which connects to $T_{\chi}^{\rm max}$ at high $T$ \cite{Knafo2019,Knafo2021,Miyake2019}.

On approaching $H_{\rm m}$, an enhancement of effective mass was observed through coefficient of the $T^2$-term of the resistivity $A$ \cite{Knafo2019,Knafo2021} and the electronic specific heat coefficient $\gamma$ derived by Maxwell's relation using magnetization data \cite{Miyake2019}.
These enhancements were discussed theoretically in terms of phenomenological Landau theory by taking into consideration of the FM fluctuation \cite{Miyake2021}.
This mass enhancement was directly confirmed by the specific heat measurements in pulsed-magnetic field and considered as an origin of the RSC transition below $H_{\rm m}$ due to the strengthened orbital limitation \cite{Imajo2019}. 
Similar mass enhancement scenario was also discussed for URhGe on the basis of a spin-triplet pairing without the Pauli limitation on both side of $H_{\rm m}$ or $H_{\rm R}$, at which the spin-reorientation occurs \cite{Miyake2008,Miyake2009}.

In contrast to the case of URhGe, the relation between MMT and SC transitions in UTe$_2$, however, is enigmatic.
For URhGe, MMT is a spin reorientation from the easy $c$ axis to the $H$ direction \cite{Levy2007}, where the development of FM fluctuation was observed \cite{Tokunaga2015}. 
Therefore, it makes a crucial role to the RSC transition, as confirmed by the field-angular \cite{Levy2007, Aoki2011s}, hydrostatic pressure \cite{Miyake2009}, and uniaxial stress dependences \cite{Braithwaite2017}.
In stark contrast, for UTe$_2$ $H_{\rm m}$ shifts to the higher $H$, but the RSC phase disappears by tilting the field directions from the $b$ to both the $a$ and $c$ axes \cite{Knebel2019, Ran2019b}.
Most spectacularly, another SC phase reemerges above $H_m$ when the $H$ is applied along certain directions between the $b$ and $c$ axes, which is near the orthorhombic [011] axis \cite{Ran2019b, Knafo2021}.
Thus, the field-orientation is a key parameter which triggers and suppresses the RSC phase around $H_{\rm m}$.

In our previous magnetization measurements, we obtained the $\gamma(H)$ using a thermodynamic Maxwell's relation assuming the isothermal magnetization processes \cite{Miyake2019}.
Owing to the significant magneto-caloric effects (MCE) \cite{Imajo2019}, this assumption was not very precise in order to discuss change in mass and the resultant RSC transition.
It is crucial to determine the field dependent effective mass more precisely. 
In particular, it is not trivial how the effective mass changes across the first-order MMT.
Here, we examine the mass enhancement more rigorously by the newly developed simultaneous measurements of magnetization and MCE with tuning the field angle.
The derived discontinuous negative (positive) jump in $\gamma(H)$ across $H_{\rm m}$ for $H~||~b$ (near [011] direction) and its influence to the disappearance (reemergence) of the RSC transition are pointed out.

Single crystals of UTe$_2$ were grown using the chemical vapor transport method \cite{Aoki2019}.
Magnetization in pulsed-magnetic fields was measured by the conventional induction method, employing coaxial pick-up coils.
To evaluate the temperature change during the field scan, we measured sample temperature simultaneously with magnetization, using a nonmagnetic ferroelectric material KTa$_{1-x}$Nb$_x$O$_3$ (KTN) as a capacitance thermometer \cite{Miyake2020}.
The measurements were performed under $^4$He gas/liquid environment.
The samples used in the simultaneous measurements of MCE and magnetization have typical dimensions with approximately 2.5 ($b$-axis) $\times$ 1.0 ($a$-axis) $\times$ 0.5 ($c$-axis)~mm$^3$.
Field direction was determined by measuring the angle between the sample and a sample holder through a microscope.
We also measured the MCE in nearly adiabatic conditions to discuss the discontinuity of the $\gamma$ at $H_{\rm m}$.
Pulsed-magnetic fields up to 56~T were generated by using non-destructive pulse-magnets having typical durations of $\sim 36$~ms, installed at the International MegaGauss Science Laboratory in the Institute for Solid State Physics of the University of Tokyo.
The field is applied along the $b$- and near [011]-axes at low temperatures down to 1.4~K.
Below 7~T and down to 1.8~K, temperature dependence of magnetization was measured by a commercial SQUID magnetometer.

Figure \ref{MHMT}(a) shows temperature dependence of magnetic susceptibility $M/H$ at $\mu_0H$~=~1~T along the $a$, $b$, $c$, and [011] axes.
The [011] axis corresponds to the field direction tilted by $\theta$~=~23.7$^{\circ}$, where $\theta$ denotes an angle measured from the $b$ to $c$ axes in the ($b$, $c$) plane.
As already known, a characteristic broad maximum appears at $T_{\chi}^{\rm max}\sim$36~K for $H~||~b$\cite{Miyake2019,Ran2019,Ikeda2006,Knafo2021}.
A similar maximum appears for $H~||~[011]$, but $T_{\chi}^{\rm max}=$~33.8~K is slightly lower than that for $H~||~b$.

In addition to the reported magnetization curves for $H~||~b$ and $a$ axes \cite{Miyake2019}, the $M(H)$ curves along the $c$ and [011] directions 
at 1.4~K are shown in Fig.~\ref{MHMT}(b).
For $H~||~c$, magnetization monotonically increases with increasing fields.
Similar to the $b$ direction, the [011]-direction magnetization shows a MMT at $\mu_0H_{\rm m}\sim$ 38~T.
$H_{\rm m}$ is consistent with the reported value, which follows the $1/\cos{\theta}$ dependence \cite{Ran2019b}. 
The low-field magnetization slope increases with increasing $\theta$, owing to the additional contribution from the $c$-axis component. 
Interestingly, an amplitude of the discontinuous jump in magnetization $\Delta M\sim 0.5~\mu_{\rm B}$ at $H_{\rm m}$ is almost same between $b$ and [011] direction.
Here, it is also interesting to emphasize that the slope above $H_{\rm m}$, namely the differential susceptibility $\chi_i \equiv dM/dH_i$ with $i =a, b, c$ and [011], is almost identical among $b$, $c$ and their intermediate [011] directions, while the magnetization for $a$-axis almost saturates above 40~T; more precisely, above 16~T $\chi_b$ becomes larger than $\chi_a$.
It indicates a weak magnetic anisotropy in the polarized paramagnetic (PPM) state within the ($b$, $c$) plane between $\chi_b$ and $\chi_c$.

\begin{figure}
\begin{center}
\includegraphics[width=\hsize]{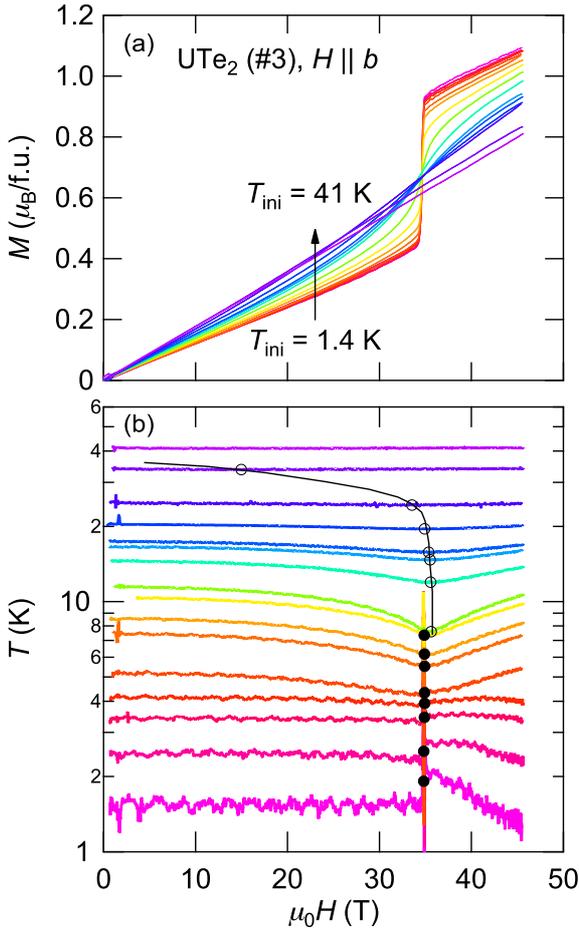}
\end{center}
\caption{(Color online)
Magnetic field dependences of (a)magnetization and (b)temperature of UTe$_2$ for $H~||~b$ axis at various initial temperatures.
For simplicity, field up-sweep curves are only shown.
The data shown in the same color in (a) and (b) were measured simultaneously.
Filled and open circles indicate ($H_{\rm m}$, $T_{\rm m}$) below and above critical end point, whose definition is discussed in ref.~\citen{suppl}.
The thin solid line on a guide of the crossover line.
}
\label{MMCE_Hb}
\end{figure}

To reveal $H$ dependence of the effective mass more precisely, we measured magnetization and MCE simultaneously.
It is noted that the samples used here are different from those used for the measurements shown in Fig.~\ref{MHMT}.
The magnetization and the sample temperature during the field up-sweep of UTe$_2$ for $H~||~b$ measured simultaneously are shown in Fig.~\ref{MMCE_Hb}.
The magnetization curves reasonably agree with our previous results \cite{Miyake2019}.
As shown in Fig.~\ref{MMCE_Hb}(b), $T(H)$ curves reveal several salient features.
First, we obtain a concrete thermodynamic evidence of first-order MMT for both $b$ and [011] axes \cite{suppl}, as reported previously \cite{Miyake2019,Knafo2019,Knafo2021,Imajo2019}.
Across $H_m$ for field-up process at the initial temperature $T_{\rm ini}=1.4$~K, $T(H)$ shows a sharp peak, followed by a positive step-like jump across $H_{\rm m}$.
The peak is most likely due to the heating effect on our KTN thermometer caused by the magnetostriction of UTe$_2$ \cite{suppl}.
The change in volume at $H_m$ will be discussed later.
The jump and peak in the $T(H)$ across $H_{\rm m}$ is also seen for the down-sweep \cite{suppl}.
Second, the mass enhancement at $H_{\rm m}$ is confirmed.
When the sample is immersed in the $^4$He gas, the temperature decreases significantly toward $H_{\rm m}$, e.g., 8.6~K at 0~T decreases to 6.2~K at $H_{\rm m}$.
In this temperature region, the measurements may be performed in more adiabatic conditions.
Assuming the isentropic measurements, i.e., $S$~=~$\int{C/T}dT$ = constant, the decrease in temperature means the increase of $\gamma$~$\sim$~$C/T$. 
Third, we could determine the CEP thermodynamically.
The peak in $T(H)$ at $H_{\rm m}$ becomes less pronounced as temperature increases, and disappears above 7.6~K \cite{suppl}, indicating that the first-order MMT becomes to crossover above the CEP.
This value is consistent with the reported phase diagram via the resistivity measurements \cite{Knafo2019, Knafo2021} but smaller than our previous magnetization results \cite{Miyake2019}.
As in our previous report not considering MCE, we overestimated the CEP temperature.

\begin{figure}
\begin{center}
\includegraphics[width=\hsize]{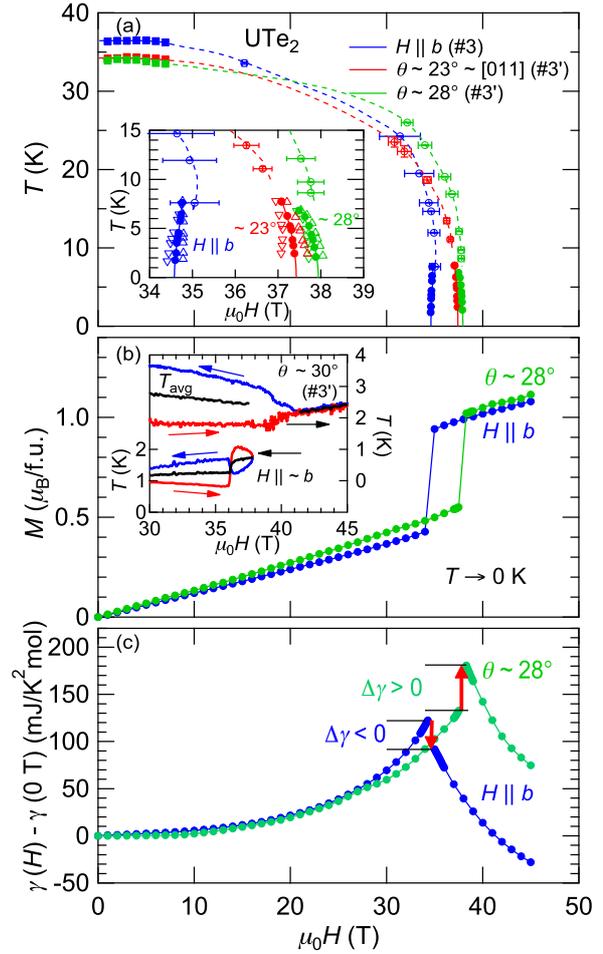}
\end{center}
\caption{(Color online)
(a) Magnetic phase diagrams of UTe$_2$ for $H~||~b$ and $\theta$$\sim23^{\circ}$ and $\sim28^{\circ}$. 
The inset focuses on low temperature parts.
Open upward (downward) triangles, filled circles, open circles, filled squares indicate $H_{\rm m}$ for up (down) sweep, $H_{\rm m}$ of MMT, $H_{\rm m}$ of crossover, and $T^{\rm max}_{\chi}$.
The broken and solid lines at low temperatures are guides and the $T^2$-fit on $H_{\rm m}(T)$, respectively.
(b) Magnetization curves extrapolated to 0~K of UTe$_2$ for $H~||~b$ and $\theta\sim28^{\circ}$. 
The inset shows the field dependence of temperature measured in quasi adiabatic conditions for $H~||~b$ \cite{Imajo2019} and $\theta$~$\sim$~30$^{\circ}$. 
The red and blue curves are measured with the field up- and down-sweep processes, and the black lines are average of up and down sweeps.
(c) Magnetic field dependence of $\gamma(H)-\gamma(0)$.
As shown by arrows, $\Delta\gamma$ changes from negative to positive with $\theta$. 
}
\label{gH}
\end{figure}

Figure \ref{gH}(a) shows the magnetic phase diagram determined by the present MCE measurements for the fields tilting from the $b$ to $c$ axes in the $(b,c)$ plane \cite{suppl}.
The MMT changed from first-order transition (closed circles) to crossover (open ones) above $\sim7.6$~K for $H~||~b$.
$H_{\rm m}$ is shifted to lower fields as temperature increases, and finally merges to $T_{\chi}^{\rm max}$.  
With increasing $\theta$, $T_{\chi}^{\rm max}$ at low fields (closed squares) shifts to lower temperature, while $H_{\rm m}$ increases.

The present study have clearly revealed that $H_{\rm m}(T)$ at lower $T$ below CEP strongly depends on the field orientation tilted from the $b$ to $c$ axes.
The inset of Fig.~\ref{gH}(a) focuses $H_{\rm m}(T)$ at low temperatures.
The CEP temperature keeps nearly constant with $\theta$.  
For $H~||~b$, $H_{\rm m}$ increases with increasing temperature.
In contrast, $H_{\rm m}(T)$ near [011] direction decreases with temperature. 
This sign change of the $H_{\rm m}(T)$ with $\theta$ for UTe$_2$ seems to affect the RSC near $H_{\rm m}$.
In addition, the slope of $H_{\rm m}(T)$ curve becomes steeper for the larger $\theta$.

From the simultaneous measurements of magnetization and temperature, we can evaluate unambiguously $M(T)$ curves at constant fields \cite{suppl}. 
At low temperature, $S = \gamma T$ for UTe$_2$ \cite{Ran2019,Aoki2019}, where $S$ is entropy.
From Maxwell's relation on each side of $H_{\rm m}$ ($H > H_{\rm m}$ and $H < H_{\rm m}$), we can directly access the $H$ dependence of $\gamma$ \cite{Miyake2019},
$\left(\partial \gamma/\partial H\large\right)_T = \left(\partial^2 M/\partial T^2\right)_H$.
The Maxwell's relation, however, is not applicable at the first-order transition to obtain the field dependence of $\gamma$ crossing $H_{\rm m}$.
This $\gamma$ is expected to change discontinuously across the first-order transition at $H_{\rm m}$.
 
To reveal the discontinuous change $\Delta \gamma$ at $H_{\rm m}$, we employ Clausius-Clapeyron equation, 
$\mu_0dH_{\rm m}/dT=-\Delta S/\Delta M$,
where $\Delta S$ and $\Delta M$ are jump/drop in entropy and magnetization crossing $H_{\rm m}$.
For itinerant metamagnets, $H_{\rm m}$ at low temperatures has been discussed to increase with $T^2$ theoretically \cite{Yamada1993}.
This temperature dependence was confirmed experimentally for the Co-based Laves phase compounds \cite{Murata1993,Goto1994, Mitamura1995} and UCoAl \cite{Muschnikov1999}.
The $H_{\rm m}(T)$ curves below CEP for UTe$_2$ also proportional to $T^2$ [see the inset of Fig.~\ref{gH}(a)].
Importantly, this $T^2$-dependence satisfies a thermodynamic principle $S$~=~0 at $T$~=~0. 
This fact also indicates that the $\Delta S$ across $H_{\rm m}$ is mainly governed by the electronic entropy, i.e., $\Delta S = \Delta \gamma T$. 
It is mentioned that an additional $T^4$-term of $\Delta S$, corresponding to the lattice contribution, reproduces the $H_{\rm m}(T)$ for $H~||~b$ much better than the only $T^2$-term.
On the other hand, the $T^2$-term is only required to describe the $H_{\rm m}(T)$ for $\theta\sim$~23 and 28$^{\circ}$.
The resultant $\Delta\gamma$ is directly obtained by $\Delta \gamma =-\mu_0\Delta M\frac{d^2H_{\rm m}}{dT^2}$.
From Fig.~\ref{gH}(a), we evaluate the coefficient of the $T^2$-term of $H_{\rm m}(T)$ as $5.3(1)\times 10^{-3}$~T/K$^2$ for $H~||~b$ direction and $-8.8(7)\times 10^{-3}$ T/K$^2$ for $\theta$~$\sim28^{\circ}$.
The magnetization curves extrapolated to 0~K shows almost same $\Delta M$ of $\sim$~0.5~$\mu_{\rm B}$ for both directions [see Fig.~\ref{gH}(b)].
Using these values, we obtain $\Delta\gamma \sim -30~(+49)$~mJ~mol$^{-1}$K$^{-2}$ for $H~||~b~(\theta$~$\sim$~28$^{\circ})$.
The reliabilities of this analysis were also reported in previous reports for some metamagnets, such as the Co-based Laves compounds ($\Delta \gamma \sim -15$~mJ/mol$^{-1}$K$^{-2}$) \cite{Murata1993,Goto1994} and UCoAl ($\Delta \gamma \sim -8$~mJ/mol$^{-1}$K$^{-2}$)\cite{Muschnikov1999}. 
As shown in the inset of Fig.~\ref{gH}(b), discontinuous increase (decrease) of $T_{\rm avg}(H)$ further supports the negative (positive) $\Delta\gamma$ (see more details in ref.~\citen{suppl}).

It should be noted that the discontinuous change of $\gamma$ at $H_{\rm m}$ is quite natural at the first-order transition. 
In the conventional first-order transition, $\Delta\gamma$ is, however, quite small as mentioned some examples previously \cite{Murata1993,Goto1994,Muschnikov1999}, because $H_{\rm m}$ is nearly constant as a function of temperature at low temperatures.
The novelty in UTe$_2$ is that $H_{\rm m}$ changes rapidly as a function of temperature with the large $\Delta M$, which gives rise to the the large jump or drop of $\gamma$ at $H_{\rm m}$; this will drastically affect to stabilize or destabilize the RSC phases.
Figure \ref{gH}(c) depicts $\gamma(H)-\gamma$(0~T) for $H~||~b$ and $\theta$~$\sim$~28$^{\circ}$.
For $H~||~b$, the obtained $\gamma(H)$ agrees rather well with specific heat results, validating our derivations \cite{Imajo2019}.
Most important finding is that the $\gamma(H)$ above $H_{\rm m}$ increases significantly for $\theta\sim28^{\circ}$, while it drops for $H~||~b$. 

Due to lack of microscopic measurements, a not-trivial question is why the rotating-field direction within the magnetic hard $(b, c)$ plane affects the slope of $H_{\rm m}(T)$. 
Similar trends of $dH_{\rm m}/dT > 0$ observed in UTe$_2$ for $H~||~b$ were also reported in UCoAl \cite{Matsuda1999,Muschnikov1999,Aoki2011B, Kimura2013} and the PM phase of UGe$_2$ under pressure \cite{Taufour2010, Kotegawa2011}, although $H$ direction is along the magnetization easy axis.
These compounds locate near the FM phase, and thus the development of the FM correlation is discussed as a main role for their metamagnetism.
As seen in UTe$_2$ for $H~||$~[011], a similar transverse-field effect, i.e., $dH_{\rm m}/dT<0$, is well established for the case of the FM URhGe \cite{Levy2005,Miyake2008,Aok12_JPSJ_review,Aoki2019re}. 
In UTe$_2$, both FM and AFM fluctuations, which are tuned by applied field, play a dominant role for the SC transitions.
Origin of MMT and its temperature dependence of $H_{\rm m}$ in UTe$_2$ may be not so simple.

Another striking point is that under pressure $H_{\rm m}$ decreases and collapses close to the pressure $P_{\rm c}$~$\sim$~1.6~GPa, where SC transition disappears in favor of a new magnetic ordered phase \cite{Braithwaite2019, Ran2020, Aoki2020, Knebel2020,Thomas2020,Aoki2021}.
Knowing the $P$ deprendence of $H_{\rm m}$ (-15~T/GPa \cite{Knebel2020}) and the jump $\Delta M$~$\sim$~0.5~$\mu_{\rm B}$, the volume shrinking at $H_{\rm m}$ is close to 0.1\%, i.e., far less than the value of $\sim$3\% which drives the system from dominant U$^{3+}$ configurations to the U$^{4+}$ ones.
At least at MMT at $H_{\rm m}$ and the valence switch at $P_{\rm c}$ the volume contracts, the link with the transition to the polarized phases deserve to be clarified.
The peak in $T(H)$ at $H_{\rm m}$ appearing below the CEP measured by our capacitance thermometer implies the volume change \cite{suppl} that was observed for the valence transition in YbInCu$_4$ \cite{Yoshimura1988}.
Moreover, this valence transition was discussed through MCE for YbInCu$_4$ \cite{Silhanek2006}, whose results share some similarities to our results in UTe$_2$.
The measurements of anisotropy of magnetostriction are in progress.

The discontinuous decrease of $\gamma$ for $H~||~b$ and increase for $\theta$~$\sim$~28$^{\circ}$ must play a key role in RSC.
If there will be only FM interactions involved, the link between the $H$ enhancement of the effective mass ($m^*$) and its feedback on SC transition can be understood as it was done for URhGe \cite{HardyPhD, Sakarya2003}, UCoGe \cite{Gasparini2010,Nikitin2017}, and other SC heavy fermion systems like UBe$_{13}$, URu$_2$Si$_2$ and UPt$_3$ \cite{Flouquet1991};
the electronic Gr\"{u}neisen parameter $\Gamma _e \equiv \frac{\partial \log m^*}{\partial \log V}$ and the SC one $\Gamma _{T_{\rm SC}} \equiv -\frac{\partial \log T_{\rm SC}}{\partial \log V}$ have opposite sign \cite{commentGru}.
In UTe$_2$, at ambient pressure $\Gamma_e$ and $\Gamma _{T_{\rm SC}}$ are both negative \cite{Thomas2021}: it is a clear mark that in UTe$_2$ an interplay between different sources of SC pairings occurs, which may be more pronounced near $H_{\rm m}$ and differ as a function of $\theta$.
In addition as the jump $\Delta M$ at $H_{\rm m}$ in UTe$_2$ overpasses that of URhGe and UCoGe by a factor 5, $H_{\rm m}$ is associated with a major Fermi surface reconstruction.
It is worthwhile to notice that despite the large jump $\Delta M$, the differential susceptibility $dM/dH$ for $H~||~b$ and $\theta$~$\sim$~28$^{\circ}$ appears quasi invariant on crossing $H_{\rm m}$.
The drastic change occurs for the contribution of localized and/or itinerant nature of the 5$f$ uranium magnetism.

In summary, we revealed the magnetization curves along the $a, b, c$, and [011] axes and discussed the mass enhancement and its impact on the  SC transition above $H_{\rm m}$ through the newly developed simultaneous measurements of magnetization and magneto-caloric effect.
Almost identical magnetization slopes above $H_{\rm m}$, indicative of the weak magnetic anisotropy, was obtained within $(b, c)$ plane, in contrast to the saturating behavior along the easy magnetic $a$ axis at low fields.
The remarkable finding is that the mass drops and jumps across $H_{\rm m}$ for $H~||~b$ and $\sim[011]$ axes.  
It is a key information to explain the experimental facts that SC transition above $H_{\rm m}$ is suppressed for the $b$ axis, but reemergenced for the [011] axis.
The main SC pairing mechanism above $H_{\rm m}$ remains unclear; an evidence of another SC paring channel different from the ones below $H_{\rm m}$ is an appealing microscopic challenge.
Finally, let us note that the field-induced RSC is detected for $H~||~c$ just above $P_c$ \cite{Aoki2021}; in this high pressure regime, a magnetic component along the $c$ axis is obviously a main actor in the interplay between the drop of $\gamma$ on crossing $H_{\rm m}$ for $H~||~b$ and an extra jump.

\begin{acknowledgment}
We thank S. Imajo, Y. Kohama, T. Kihara, T. Sakakibara, K. Miyake, K. Machida, F. Hardy, W. Knafo, G. Knebel, and J. P. Brison for fruitful discussion.
This work was supported by KAKENHI (JP15H05884, JP15H05882, JP15K21732, JP16H04006, JP15H05745, JP19H00646, JP20K03854, JP20K20889, JP20H00130, and JP20KK0061), the Precise Measurement Technology Promotion Foundation (PMTP-F), ICC-IMR, and ERC starting grant (NewHeavyFermion).
\end{acknowledgment}

\vspace{20pt}

{\bf Supplemental material}
\section{Temperature dependence of magnetization at constant fields}

\begin{figure}
\begin{center}
\includegraphics[width=0.6\hsize]{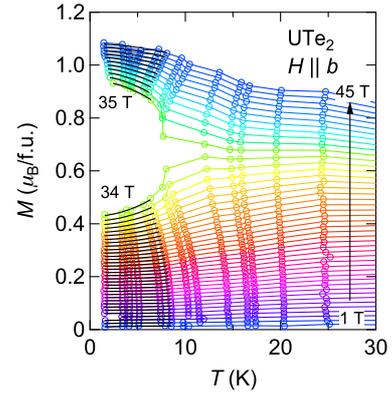}
\end{center}
\caption{Temperature dependence of magnetization of UTe$_2$ for $H~||~b$ at constant fields depicted from Fig.~2  in the main text.
Field varies by 1-T step from bottom to top.
The black solid lines are $T^2$-fit of $M(T)$ curves.
}
\label{MT_Hb}
\end{figure}

\begin{figure}
\begin{center}
\includegraphics[width=0.6\hsize]{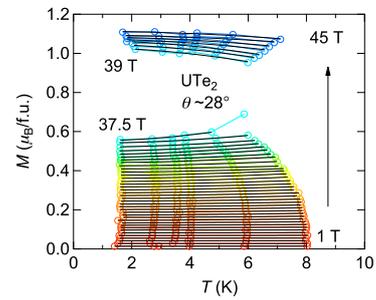}
\end{center}
\caption{Temperature dependence of magnetization of UTe$_2$ for $\theta\sim28~^{\circ}$ at constant fields below $T_{\rm ini}$~=~8~K.
Field varies by 1-T step from bottom to top except for near $\mu_0H_{\rm m}$~=~38~T.
The black solid lines are $T^2$-fit of $M(T)$ curves.
}
\label{MT_28}
\end{figure}

From the simultaneous measurements, we can evaluate $M(T)$ curves at constant fields, as shown in Figs.~\ref{MT_Hb} and \ref{MT_28}. 
For $H~||~b$, the $M(T)$ slope changes its sign from positive to negative across $H_{\rm m}$, in consistent with previous measurements \cite{Miyake2019_2}.
The importance of the simultaneous evaluations of magnetization and temperature appears near $\mu_0H_{\rm m}$~= 34.6~T: temperature decreases toward $H_{\rm m}$ in the more adiabatic conditions above 4.2~K.
Similar trend is also seen in $M(T)$ curve for $\theta\sim28^{\circ}$ (Fig.~\ref{MT_28}).
The solid lines are fitting results of the $T^2$-dependence of $M$ in Figs.~\ref{MT_Hb} and \ref{MT_28}, which reproduce well the experimental results.

\section{Anomalies in $T(H)$ curves and the definition of $(H_{\rm m},T_{\rm m})$ through the $T(H)$ curves}

\begin{figure}
\begin{center}
\includegraphics[width=0.6\hsize]{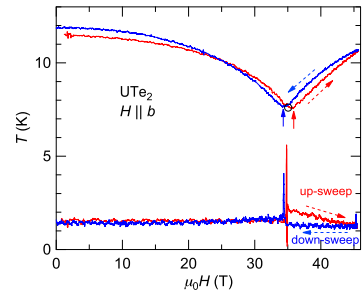}
\end{center}
\caption{Magneto-caloric effect of UTe$_2$ measured at $T_{\rm ini}=1.4$ and 11.4~K. 
}
\label{HmTm}
\end{figure}

\begin{figure}
\begin{center}
\includegraphics[width=\hsize]{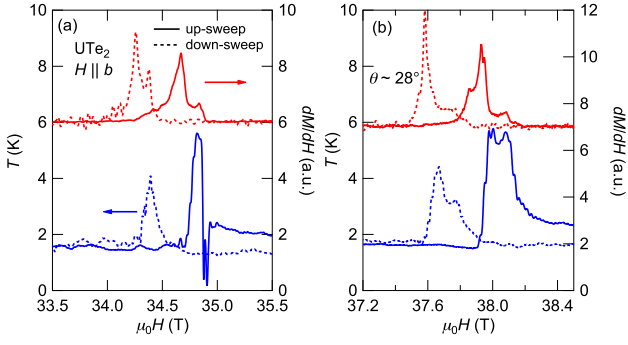}
\end{center}
\caption{Comparison of MCE and $dM/dH$ of UTe$_2$ for (a) $H~||~b$ and (b) $\theta\sim28~^{\circ}$ at lowest measured temperature.
The red and blue curves correspond to MCE and $dM/dH$, respectively.
The solid and dashed lines are obtained for the field up- and down-sweep processes.
}
\label{MCE_dMdH}
\end{figure}

\begin{figure}
\begin{center}
\includegraphics[width=\hsize]{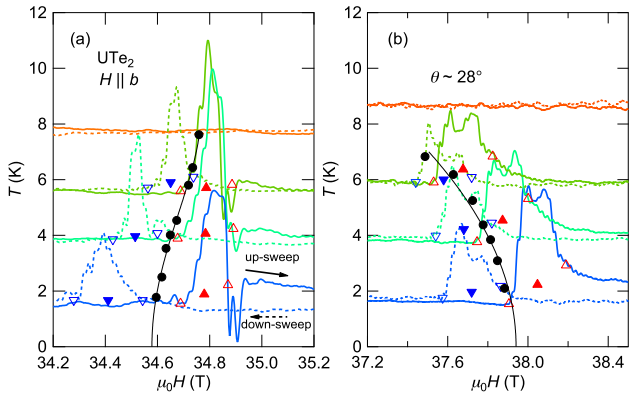}
\end{center}
\caption{The MCE curves of UTe$_2$ for $H~||~b$ and $\theta\sim28~^{\circ}$ at selected temperatures.
The solid and dashed lines are obtained for the field up- and down-sweep processes.
The triangles superimposed in the $T(H)$ curves correspond to the transition points.
The full circles point out the thermodynamic transition points.
The points, whose $T(H)$ curves are not shown, are also shown.  
See text for the detail of definition of $(H_{\rm m}, T_{\rm m})$. 
}
\label{MCE}
\end{figure}

Figure~\ref{HmTm} shows MCE results, i.e., field dependence of temperature, of UTe$_2$ measured at the initial temperature of $T_{\rm ini}=1.4$~K (in superfluid of $^4$He) and 11.4~K (in $^4$He gas) for $H~||~b$. 
A peak and a step-like jump in the $T(H)$ across $H_{\rm m}$
appear for $T_{\rm ini}=1.4$~K [see also Figs.~\ref{MCE_dMdH}(a) and \ref{MCE}(a)].
For the $H$-ascending process beyond $H_{\rm m}$, a significant decrease in $T$ is observed.
This is due to relaxation of the sample temperature to surrounding of the superfluid of $^4$He.
$T$ remains almost constant down to $H_{\rm m}$, followed by a peak and a step increment for the $H$-descending process.
Below $H_{\rm m}$, the $T(H)$ is almost identical between up and down field-sweeps, indicating that the measurements were performed in the isothermal conditions.
The peak structure in $T(H)$ curves at $H_{\rm m}$ is an intriguing phenomenon.
In our previous MCE measurements using a resistive thermometer, such a peak was not observed \cite{Imajo2019_2}.
To check the reliability of our MCE measurements by the ferroelectric capacitance thermometer, we also measure the field-induced spin reorientation in URhGe using the same capacitance thermometer.
In contrast to the peak structure in UTe$_2$, a minimum in $T(H)$ crossing a (weak) first order metamagnetic transition was only observed in the $T(H)$ curves for URhGe \cite{Miyake2021_2}.
For UTe$_2$, this peak structure is probably due to the piezoelectric effect on the thermometer and/or heating effect on the interface between the sample and the thermometer.
Magnetostriction of UTe$_2$ at $H_{\rm m}$ $\sim$~0.1\% may be much larger than that of the nonmagnetic ferroelectric material. 
This different magnetostriction leads to friction heat on the interface.
Since the thermometer has a much smaller volume and thus heat capacity than the sample, the thermometer temperature rapidly changes due to the friction heat.
Therefore, the peak in $T(H)$ is a clear indicative of the first-order nature MMT below CEP.
At $T_{\rm ini} = 11.4$~K above CEP, a minimum in $T(H)$ only appears as in Fig.~\ref{HmTm}. 
This anomaly corresponds to a crossover changed from first-order transition.   

Next, we carefully look at the MCE and magnetization results for determining a precise phase diagram.
Figure~\ref{MCE_dMdH} presents the field dependence of temperature (blue lines) and the differential susceptibility $dM/dH$ (red lines) at the lowest measured temperature for (a)$H~||~b$ and (b) $\theta\sim$28$^{\circ}$.   
For $H~||~b$, $dM/dH$ peaks around 34.7~T for field up-sweep, where the temperature starts increasing.
The $T(H)$ shows a peak, followed by a step-like jump.
The field of the $T(H)$ step coincides with the disappearance of the $dM/dH$ anomaly.
Sharp negative peak in the $T(H)$ curve just above the positive peak is an artifact.
Since the each data point is obtained by an average of one period of the sinusoidal signal with a measuring frequency of 50~kHz, our sampling rate of 1~MHz used here is not large enough to capture the sudden change \cite{Miyake2020_2}.
For field down-sweep, similarly, temperature starts increases near 34.6~T and shows a sharp peak, followed by a step-like increase at 34.2~T, where the anomaly in $dM/dH$ disappears.
These trends are also similar to the case for $\theta\sim$~28$^{\circ}$ [Fig.~\ref{MCE_dMdH}(b)]. 

We define the transition point ($H_{\rm m}, T_{\rm m}$) as follows.
For the field-up sweep, the beginning and ending points of the first order metamagnetic transition are determined at	 the foot of the peak, as pointed by open upward triangles (see Fig.~\red{\ref{MCE}}).
The ($H_{\rm m}, T_{\rm m}$) for the up-sweep is assumed as an average of the beginning and ending points (full upward triangles). 
We define ($H_{\rm m}, T_{\rm m}$) for the down-sweep by the same way (downward triangles).
Finally, the thermodynamic ($H_{\rm m}, T_{\rm m}$) is defined as an average of up and down sweeps (filled circle).
At higher temperatures than $\sim8$~K, the peak structure in the $T(H)$ curves disappears, indicating that the first order transition becomes a crossover.
As shown in Fig.~\ref{HmTm}, a minimum of $T(H)$ corresponds to the crossover point ($H_{\rm m}, T_{\rm m}$).
In addition, the $H$-hysteresis is observed in $T(H)$ curves.
This is caused by heat leak to surroundings.
Because the sample immersed in $^4$He gas atmosphere above 4.2 K, quasi-adiabatic conditions may be realized.
The $H_{\rm m}$ (pointed by a red arrow) for the up-sweep is higher than that for the down-sweep (blue arrow).
This means that there is some delay between temperatures of thermometer and sample (see ref.~\citen{Kihara2013}).
We define ($H_{\rm m}, T_{\rm m}$) above the CEP as an average of the minimum of $T(H)$ curves for up- and down-sweep, which is shown as horizontal error bars in Fig.~3 of the main text.

\section{Field-angular dependence of the magnetic phase diagram}

\begin{figure}
\begin{center}
\includegraphics[width=\hsize]{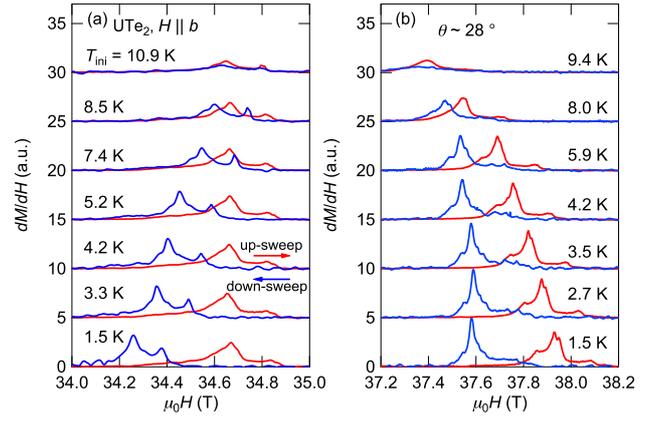}
\end{center}
\caption{Differential susceptibility of UTe$_2$ for (a) $H~||~b$ and (b) $\theta\sim28~^{\circ}$ at several initial temperatures, $T_{\rm ini}$.
The red and blue curves are measured for the field up- and down-sweep, respectively.
}
\label{dMdH}
\end{figure}

Key observation in this work is that the change of the slope in the $H_{\rm m}(T)$ curves, by tilting the field direction from the $b$ to $c$ axis.
One can clearly see differences of $H_m(T)$ between $H~||~b$ and $\theta~\sim$~28$^{\circ}$ through the temperature dependence of differential susceptibility $dM/dH$ shown in Fig.~\ref{dMdH}. 
For $H~||~b$, sharp anomalies corresponding to the metamagnetic transition shift to the higher fields with increasing $T_{\rm ini}$.
On the other hand, for $\theta~\sim$~28$^{\circ}$ the anomalies in $dM/dH$ move to the lower fields as $T_{\rm ini}$ increases.
Here, it is noted that the temperatures labeled in Fig.~\ref{dMdH} do not correspond to the temperatures at $H_{\rm m}$ due to the MCE.
Therefore, we employ the MCE data for the precise determination of phase diagram.

As already discussed the definition of $(H_{\rm m}, T_{\rm m})$ in the previous section, we determine the phase diagram through the MCE results shown in Fig.~\ref{MCE}.
As pointed out by the black circles, $H_{\rm m}(T)$ shows the positive $T$ dependence for $H~||~b$, while the negative-dependence for $\theta~\sim$~28~$^{\circ}$. 

\section{Discontinuous change in $\gamma(H)$ at $H_{\rm m}$}
One of the remarkable observation is change in the sign of the discontinuous jump of $\gamma(H)$, $\Delta \gamma$, at $H_{\rm m}$ with tilting the field direction from the $b$ to $c$ axis.
The $\Delta S$ across $H_{\rm m}$ is negative (positive) for $H~||~b$ ($\theta\sim28^{\circ}$).
Since we evaluate the $\Delta \gamma$ using the thermodynamic relation, namely Clausius-Clapeyron equation, one may argue reliability of this analysis.
To show the reliability, we performed the almost adiabatic 
MCE measurements for the $H$-direction near the [011] axis.
The adiabatic MCE occurs in the isentropic conditions.
The change in sample temperature reflects the $\gamma(H)$ in ideal adiabatic conditions.

\begin{figure}
\begin{center}
\includegraphics[width=\hsize]{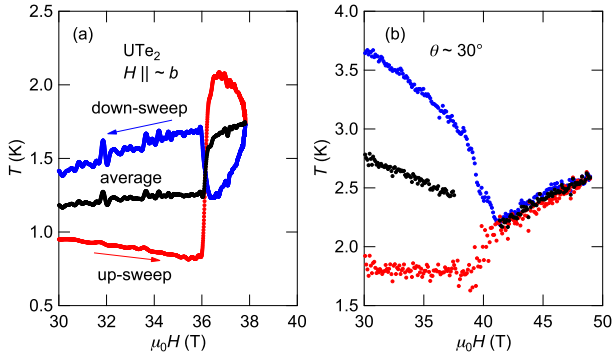}
\end{center}
\caption{MCE of UTe$_2$ for (a) $H~||~\sim b$\cite{Imajo2019} and (b) $\theta\sim30~^{\circ}$ at the lowest-measured initial temperature.
The red and blue symbols are measured for the field up- and down-sweep, respectively.
The black symbols correspond to the average of the MCE curves for the field up- and down-sweep.
}
\label{MCE_comp}
\end{figure}

\begin{figure}
\begin{center}
\includegraphics[width=0.6\hsize]{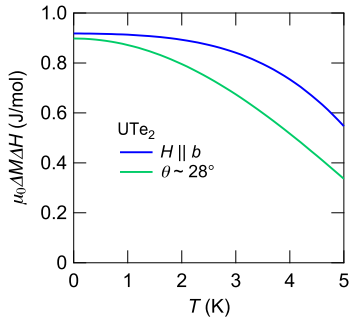}
\end{center}
\caption{Temperature dependence of hysteresis loss $\Delta\mu_0\Delta H$ of UTe$_2$ for $H~||~b$ and $\theta$~$\sim$28$^{\circ}$.
}
\label{Loss}
\end{figure}

Figure~\ref{MCE_comp} shows the MCE for $H~||~b$ and $\theta$~$\sim$~30$^{\circ}$ measured in the nearly adiabatic conditions.
For $H~||~b$, the data is taken from the previous report \cite{Imajo2019_2}.
In order to make the adiabatic condition for the measurements for $\theta$~$\sim$~30$^{\circ}$, the sample space is evacuated.
For both up and down field sweep, $T(H)$ increases across $H_{\rm m}$ due to hysteresis loss.
Since the MMT of UTe$_2$ accompanies a large $\Delta M$ with the $H$ hysteresis, the loss ($\sim\Delta M \Delta H$) is significant. 
We average $T(H)$ of the up-sweep and down-sweep to remove the contribution of the loss.
It is noted that the loss varies as a function of temperature.
In general, the loss increases with decreasing temperature.
Since the sample temperatures at $H_m$ are different between the up- and down-sweep measurements, the averaging of $T(H)$ may not be a rigorous procedure to remove the loss contribution from the $T(H)$ curves.
To check the reliability of this average procedure, we estimated temperature dependence of $\Delta M\Delta H$ using the $M(T)$ and $H_{\rm m}(T)$ curves, as shown in Fig.~\ref{Loss}.
As expected, the $\Delta M\Delta H$ increases with decreasing temperature.
For $\theta$~$\sim$~28$^{\circ}$, however, the increment of $\Delta M\Delta H$ becomes smaller as $T$ lowers.
Assuming the same temperature dependence of $\Delta M\Delta H$, for $\theta$~$\sim$~30$^{\circ}$ only $\sim$5\% larger for the up-sweep ($\sim$1.8~K) than that for the down-sweep ($\sim$2.2~K) [See Fig.~\ref{MCE_comp}(b)].
Moreover, temperature variation of $\Delta M\Delta H$ for $H~||~b$ is nearly constant for the experimental temperature regions [See Fig.~\ref{MCE_comp}(a)].
From these results, the simple averaging procedure can satisfactorily remove the loss contribution.

As clearly seen, the average MCE, $T_{\rm avg}(H)$ just above $H_{\rm m}$ shows difference between $H~||~b$ and $\theta\sim30^{\circ}$.
$T_{\rm avg}(H)$ increases discontinuously for $H~||~b$.
In contrast, $T_{\rm avg}(H)$ for $\theta\sim30^{\circ}$ decreases discontinuously, although the amplitude of the discontinuity is weaker than that for $H~||~b$ due to the rather high temperature measurements, i.e., the heat capacity of sample is higher, and thus the temperature change is smaller.  
These jump and drop of $T_{\rm avg}(H)$ curves support our estimation of the negative and positive $\Delta\gamma$.
From these experimental observation, we conclude that our estimation of $\Delta\gamma$ using the thermodynamics are trustworthy.
For quantitative analysis, we need the MCE measurements with the exactly same experimental conditions.
It is also interesting to investigate the field-sweep-rate dependence of $T(H)$, that is left for further studies \cite{KMiyake2021}.

Finally, we comment about a difference between Fig.~\ref{MCE}(b) and Fig.~\ref{MCE_comp}; $T(H)$ shows a peak near $H_{\rm m}$ for $\theta\sim 28^{\circ}$ [Fig.~\ref{MCE}(b)], whereas the peak disappears for $\theta\sim 30^{\circ}$ (Fig.~\ref{MCE_comp}), although the sample and thermometer used are same.
We performed the measurements for $\theta\sim 30^{\circ}$ with different setup form that for $\theta\sim 28^{\circ}$.
Considering almost identical $T(H)$ curves above $H_m$ between up- and down-sweep (Fig. 7), sample was in nearly adiabatic conditions, and the thermal coupling between sample and thermometer is adequate. 
It is not clear why the peak disappears for $\theta\sim 30^{\circ}$ at present. 
Several reasons of this discrepancy are considerable. 
First, sample may be damaged through the magnetostriction across $H_{\rm m}$. 
As mentioned in Ref.~\citen{Miyake2019_2}, amplitude of magnetization jump accompanied by the metamagnetic transition decreases slightly after several thermal and field cycles. 
Second, the field-angle may affect the amplitude of the peak. 
The data shown in Fig.~\ref{MCE_comp} is obtained for the larger $\theta$ than that shown in Fig.~\ref{MCE}(b). 
Third, the first-order transition may depend on environmental conditions; adiabatic or isothermal conditions affect the $T(H)$ curves.
To discuss the peak structure more precisely, we need to perform the MCE measurements with keeping the same experimental setup by changing the sample environments. 
These problems are left for future studies.


\begin{thebibliography}{99}
\bibitem{commentGru} Experimentally, Gr\"{u}neisen parameter is obtained by $\Gamma = \frac{V_m\alpha_V}{\kappa C}$, where $V_m$, $\alpha_V$, $\kappa$, and $C$ are molar volume, volume thermal expansion coefficient, compressibility, and specific heat.
The sign of $\Gamma$ depends on $\alpha_V$.
\end{thebibliography}
\end{document}